%% file: poster_hawaii.tex
\documentclass[twoside]{reportj}  
\usepackage{color}
\usepackage{graphicx}
\input{isolatin.sty}
\input{jesus}

\def\pio{\mbox{$\pi_{\rm o}$}}
\def\spi{\mbox{$\sigma_\pi$}}
\def\zsun{\mbox{$z_\odot$}}
\begin{document}
\definecolor{Black}{rgb}{0.0,0.0,0.0}
\definecolor{White}{rgb}{1.0,1.0,1.0}

\centerline{\textcolor[rgb]{0.8,0.0,0.0}{\fontsize{0.28in}{0.32in}\selectfont 
\bf Accurate distances to nearby massive stars with}}
\medskip
\centerline{\textcolor[rgb]{0.8,0.0,0.0}{\fontsize{0.28in}{0.32in}\selectfont 
\bf the new reduction of the Hipparcos raw data}}
\bigskip
\centerline{\textcolor[rgb]{0.0,0.5,0.0}{\fontsize{0.14in}{0.15in}\selectfont 
\bf Jesús Maíz Apellániz (IAA-CSIC), Emilio J. Alfaro (IAA-CSIC), \& Alfredo Sota (STScI)}}
\bigskip
\centerline{\textcolor[rgb]{0.0,0.0,0.8}{\fontsize{0.14in}{0.15in}\selectfont 
\bf Poster presented at IAU Symposium 250: Massive stars as cosmic engines, 10-14 December 2007}}
\bigskip

\section*{The new Hipparcos reduction}

\ind	van Leeuwen (2007) has produced a new reduction of the Hipparcos data that includes a careful modeling of the satellite dynamics
and eliminates the data correlations in the original catalog caused by attitude-modeling errors. As a result, the distances to the
Pleiades and other clusters are now consistent with the values obtained with other methods. The new analysis also reduces the parallax 
uncertainties for the brightest stars by up to a factor of four. We use the new results to study the spatial distribution of massive
stars in the solar neighborhood and to improve the distances to individual objects.

\section*{The technique}

\ind	For a star with an observed parallax \pio\ and Gaussian uncertainty \spi, its distance probability distribution is given by:

\begin{equation}
p(r|\pio) = A\, r^2\,{\rm exp}\left(\frac{1-r\pi_{\rm o}}{\sqrt{2}r\sigma_\pi}\right)^2 \rho(r), \label{dist}
\end{equation}

\noindent where $A$ is a normalization constant and $\rho(r)$ is the underlying spatial distribution (which can be thought of as a
Bayesian prior). Note that, in general:

\begin{equation}
<p(r|\pio)> \ne 1/\pio 
\end{equation}

\noindent because of the conversion from parallax to its inverse, the availability of larger volumes of space behind 1/\pio\ than in
front of it, and the possibility of a non-constant $\rho(r)$. This effect is known as the Lutz-Kelker bias (Lutz \& Kelker 1973), who
described it for the special case of constant $\rho(r)$ (Figure 1). If $\rho(r)$ is well-behaved (i.e. it has a cutoff for large $r$) and known, 
one can apply a Lutz-Kelker correction $c$ such that:

\begin{equation}
d = 1/\pio + c,
\end{equation}

\noindent where $d$ is the median distance of $p(r|\pio)$. Note that for a constant $\rho(r)$, $c$ is strictly infinite but can be made 
finite by introducing an artificial distance cutoff if $\spi/\pio < 0.175$. $c$ is positive in most cases, with the exceptions arising when $\rho(r)$ has
a strong negative slope.

	In many astrophysical cases of interest $\rho(r)$ is unknown a priori (e.g. one is measuring parallaxes precisely to obtain it).
Under those circumstances, it may be possible to use the method described by Maíz Apellániz (2001,2005). One starts by selecting a large homogeneous
sample of objects with measured parallaxes and assumes a parameterized $\rho(r)$, with the initial values for the parameters chosen from the 
literature. $p(r|\pio)$ is then calculated for each star using Eqn.~\ref{dist} and the results are used to build a new $\rho(r)$ with a 
$\chi^2$ minimization algorithm that applies the binning recipe of Maíz Apellániz \& Úbeda (2005). The procedure is then iterated until 
the parameters converge. At the end of the process, one can check whether the choice for the $\rho(r)$ parameterization was good and redo the 
full procedure with a new one if necessary.

\centerline{\includegraphics*[width=0.47\linewidth]{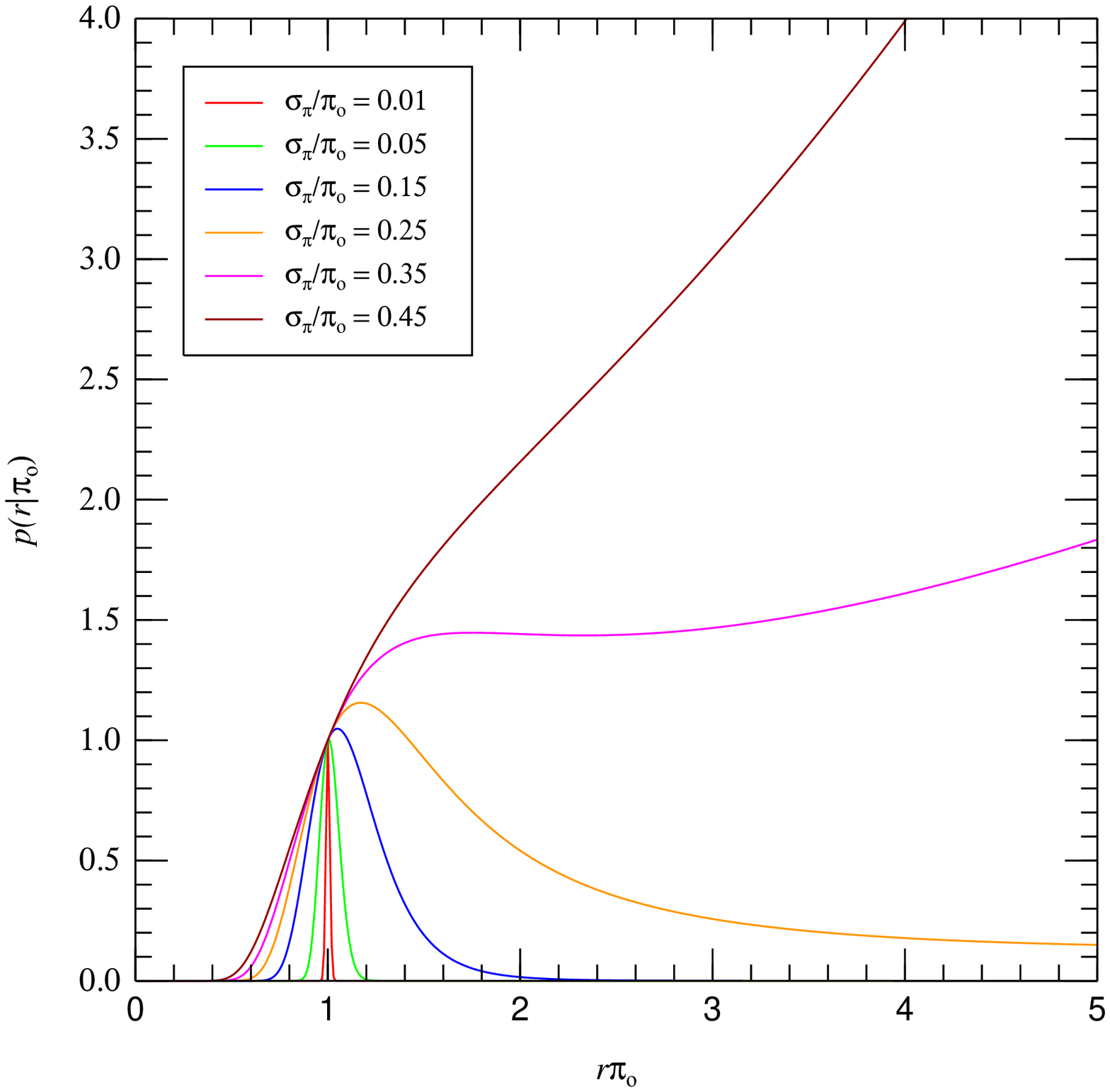} \
            \includegraphics*[width=0.47\linewidth]{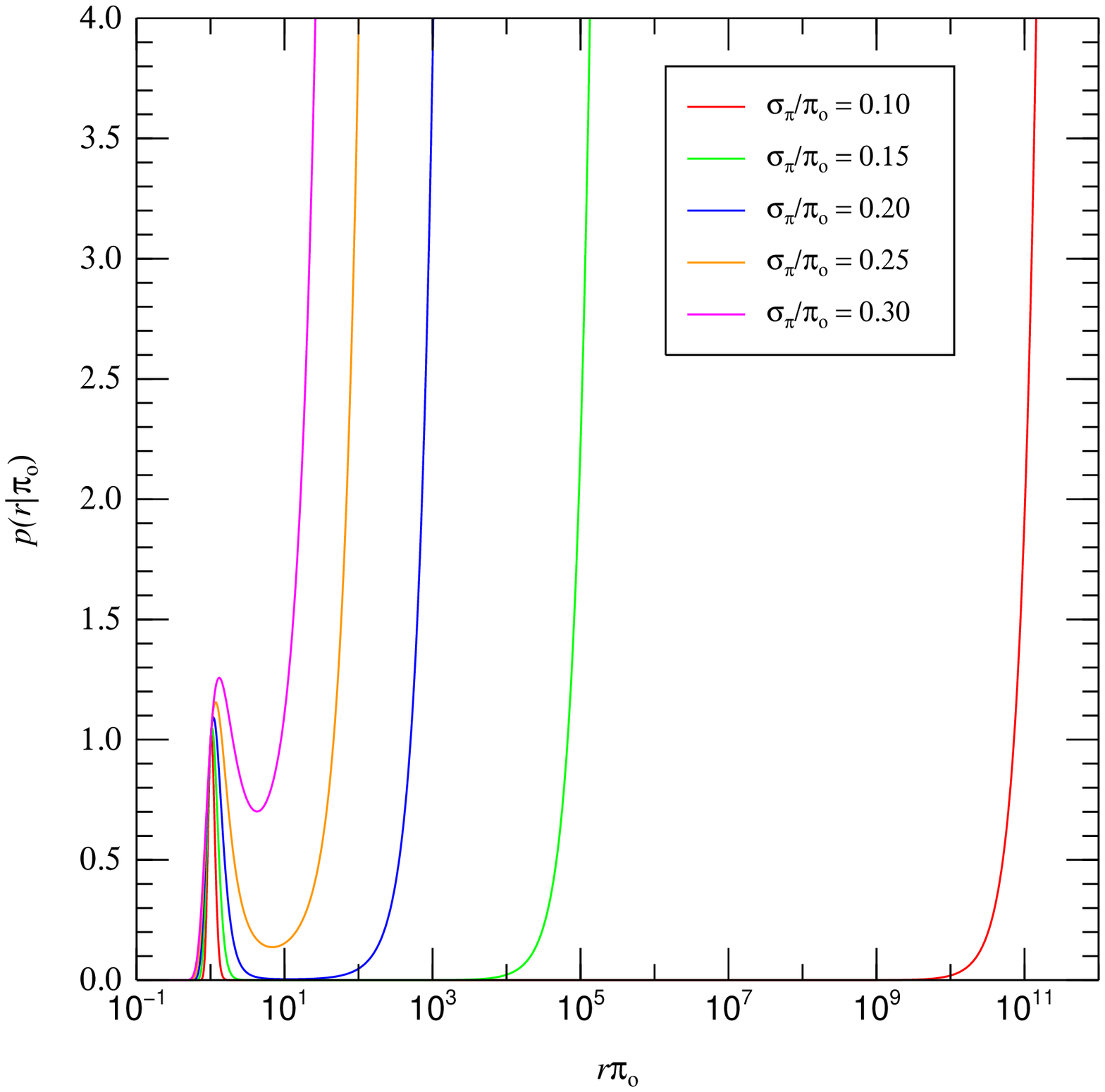}}

\noindent {\bf Figure 1.} $p(r|\pio)$ for a constant $\rho(r)$ and different values of \spi/\pio\ normalized to the value at $r$ = 1/\pio. 
The left panel has a linear horizontal scale and the right 
one a logarithmic one. Note how in the classical Lutz-Kelker case $p(r|\pio)$ always diverges for large $r$.

\bigskip

\centerline{\includegraphics*[width=1.0\linewidth]{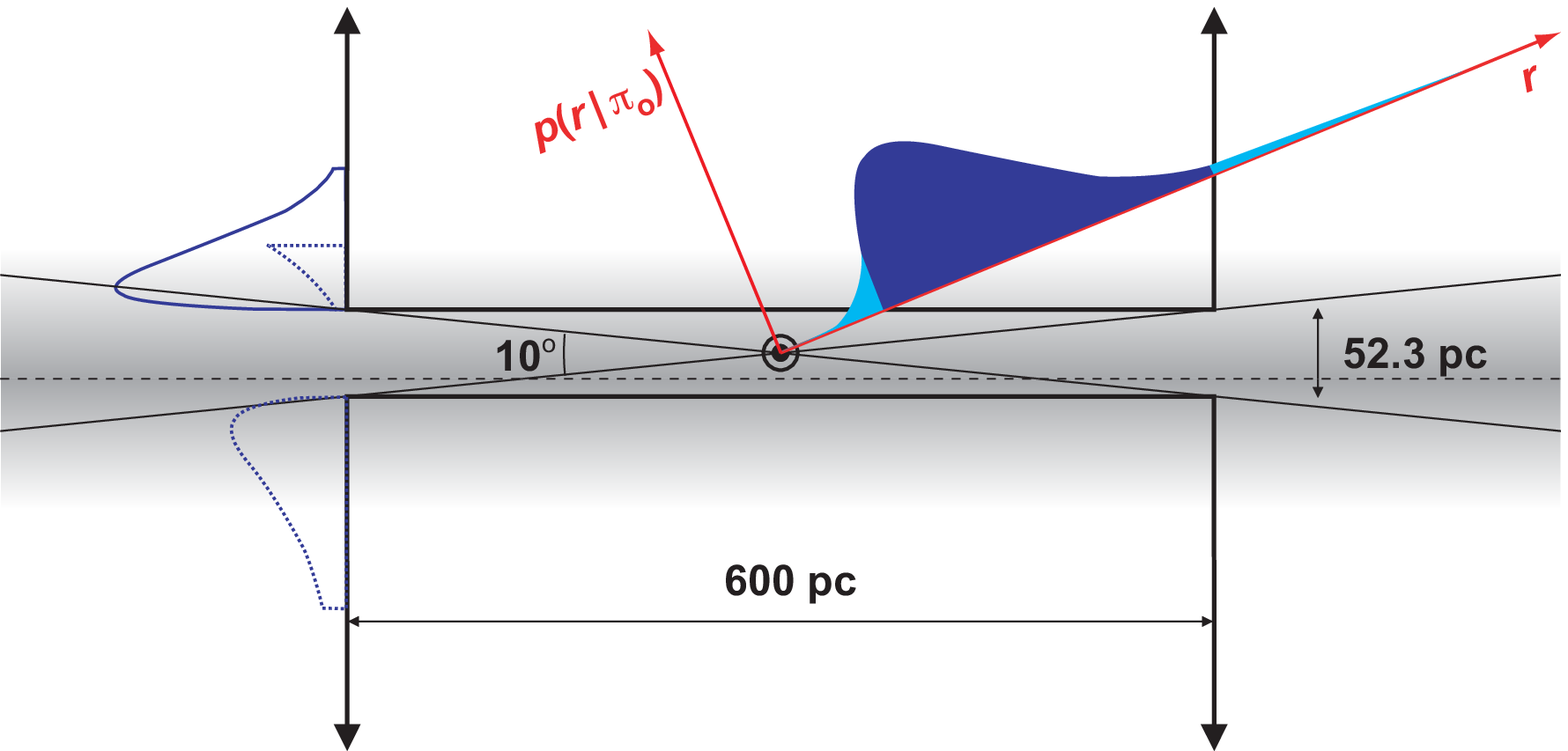}}

\noindent {\bf Figure 2.} Galactic vertical structure model and double semi-infinite cylinder used. $\odot$ marks the Sun position 
and the horizontal dashed line the Galactic plane. A sample $p(r|\pio)$ (filled curve) and its projection on the vertical axis inside the considered volume 
(blue continuous line) are shown. Two additional sample projections of $p(r|\pio)$ are also shown (blue dashed lines).

\section*{Sample and parameterization}

\ind	At low Galactic latitudes $b$, the Hipparcos data are complete only for $m_V \le 7.9$. Therefore, we can only study the massive stars
in the solar vicinity. Using the data in the Hipparcos catalog itself, we selected the objects with an spectral classification of WR, O, 
or B0-B5, as well as all stars of luminosity type I. We also searched additional catalogs (e.g. van der Hucht 2001, Maíz Apellániz et al. 
2004) to check for possible omissions and analyzed the cases where the derived absolute magnitude was anomalously large in order to
discard stars with erroneous spectral classifications (a total of 93). The final sample has 69 WR stars, 293 O stars without a WR
companion, 871 non-O stars with luminosity class I, and 2770 non-supergiant B stars (a total of 4003 objects).

	Following Maíz Apellániz (2001), we select a $\rho(r)$ that depends only on the vertical Galactic coordinate $z$ and that is
composed of a self-gravitating, isothermal disk plus a Gaussian halo:

\begin{equation}
\rho(z) = \rho_{d,0}\, {\rm sech}^2 \left(\frac{z+\zsun }{2h_d}\right) +
          \rho_{h,0}\, \exp\left(\frac{z+\zsun }{\sqrt{2}h_h}\right)^2 \label{dens}
\end{equation}

\noindent or, equivalently:

\begin{equation}
\rho(z) = \sigma\left(\frac{1-f}{4h_d}\, {\rm sech}^2 \left(\frac{z+\zsun }{2h_d}\right) +
                      \frac{f}{2\sqrt{2}h_h}\, \exp\left(\frac{z+\zsun }{\sqrt{2}h_h}\right)^2\right), \label{dens2}
\end{equation}

\hyphenation{res-pec-ti-ve-ly}

\noindent where $\rho_{d,0}$ and $\rho_{h,0}$ are the disk and halo volume number densities at $z=0$, respectively; $h_d$ and $h_h$
are the disk scale height and halo half width, respectively; \zsun\ is the Sun's distance above the Galactic plane; $\sigma$ is the total surface
number density; and $f$ is the fraction of stars in the halo population. In order to
minimize incompleteness effects due to extinction and distance when calculating the Galactic vertical structure parameters, we consider only 
the objects located in a double semi-infinite cylinder with a radius of 300 pc and a zone of avoidance given by $|b| = 5^{\rm o}$ (Figure~2). 

\bigskip

\centerline{\includegraphics*[width=0.5\linewidth, bb=28 28 566 540]{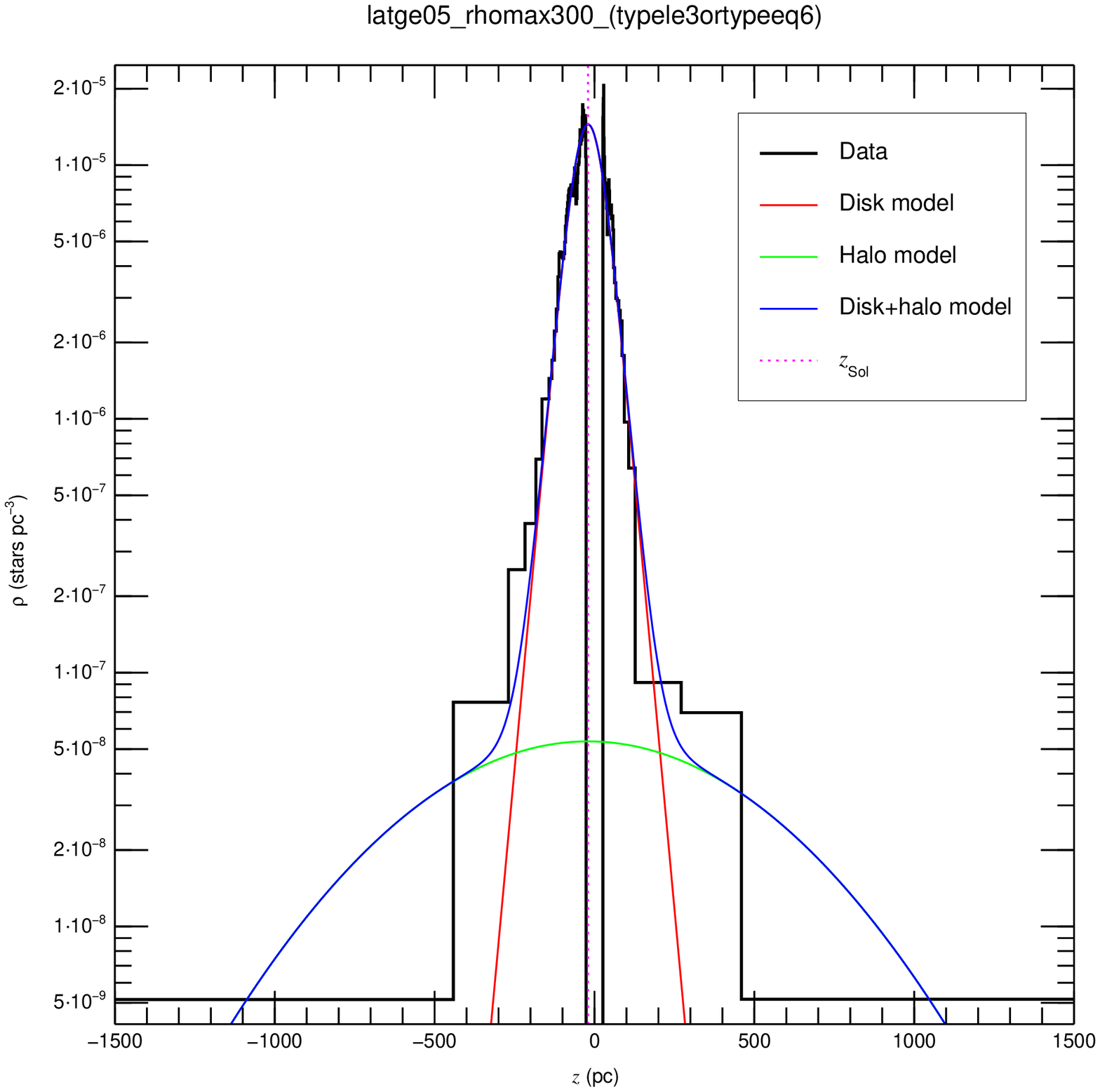}}

\noindent {\bf Figure 3.} Observed and model fit for the Galactic vertical structure for massive stars in the solar vicinity.

\centerline{\includegraphics*[width=0.6\linewidth, bb=28 28 566 540]{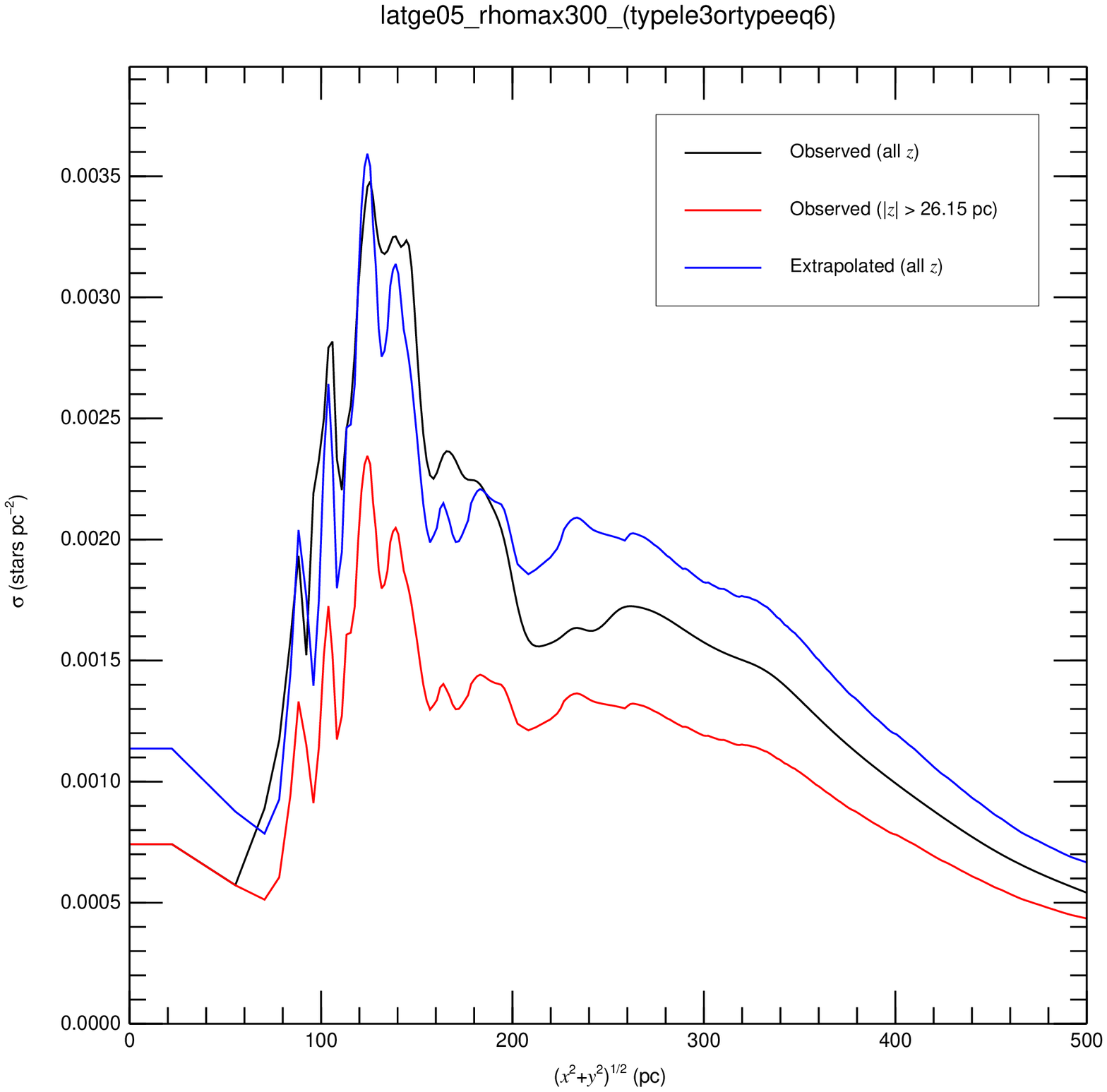}}

\noindent {\bf Figure 4.} Massive-star surface densities as a function of distance from the Sun vertical line. The black and red lines show the observed
densities for all Galactic latitudes and for the region with $|z| > 26.15$~pc, respectively. The blue line is the red line multiplied by a factor that
accounts for the stars in the gap defined by $|z| \le 26.15$~pc.

\section*{Results}

\begin{itemize}
 \item The values of $\zsun = 20.0 \pm 2.9$ pc and $h_d = 31.8 \pm 1.6$ pc (see Figure 3) are similar to the ones obtained by Maíz Apellániz (2001) and
       other authors.
 \item The results for the other three parameters are $\sigma = (1.91 \pm 0.11)\cdot 10^{-3}$~stars~pc\mm/, $h_h = 490 \pm 170$~pc, and 
       $f = 0.039\pm 0.015$. The low value of $f$ indicates that the number of runaway stars in the sample cannot be too large.
 \item There is little variation in the fitted results when selecting subsamples by spectral type or luminosity class.
 \item There are large variations in the results when fitting subsamples along different Galactic quadrants. Specifically, \zsun\ is larger for the third 
       quadrant and smaller for the fourth one. Those results are caused by the presence of local disturbances in the distribution of massive stars (the 
       Orion OB1 and the Scorpius-Centaurus OB associations, see e.g. Elías et al. 2006).
 \item There are also large variations in the massive-star surface density as a function of distance from the Sun along the Galactic plane (Figure 4). 
       As previously noted by Maíz Apellániz (2001), the Sun is placed at a local $\sigma$ minimum. A clear maximum exists at
       $100-160$ pc, mostly due to the Scorpius-Centaurus OB association. At a distance of $200$ pc, extinction starts to affect the observed population
       of massive stars near the Galactic equator and at $300-350$ pc we reach the Hipparcos completeness limit even for higher latitudes.
 \item The distance uncertainties for the nearest massive stars have been substantially reduced and the number of massive stars with significative Hipparcos 
       distances has been increased. See Tables~1~and~2 and Figure~5 for examples.
 \item The new Hipparcos distance to $\gamma^2$ Vel is in excellent agreement with the recent values of \vsmsp{336}{7}{8} pc derived with SUSI (North et al.
       2007) and of \vsmsp{368}{13}{38} pc derived with VLTI (Millour et al 2007). Note also that the old Hipparcos value was less than 2 sigmas away from 
       any of the current ones. It is often erroneously quoted that the old
       Hipparcos distance to $\gamma^2$ Vel was \vsmsp{258}{31}{41} pc but that 
       value does not include the Lutz-Kelker correction $c$.
 \item The new Hipparcos distance to $\theta^2$ Ori A is also in good agreement (1 sigma) with the VLBA distance to the Orion nebula of $414\pm 7$ pc 
       recently derived by Menten et al. (2007).
\end{itemize}

\bigskip

\centerline{\includegraphics*[width=\linewidth]{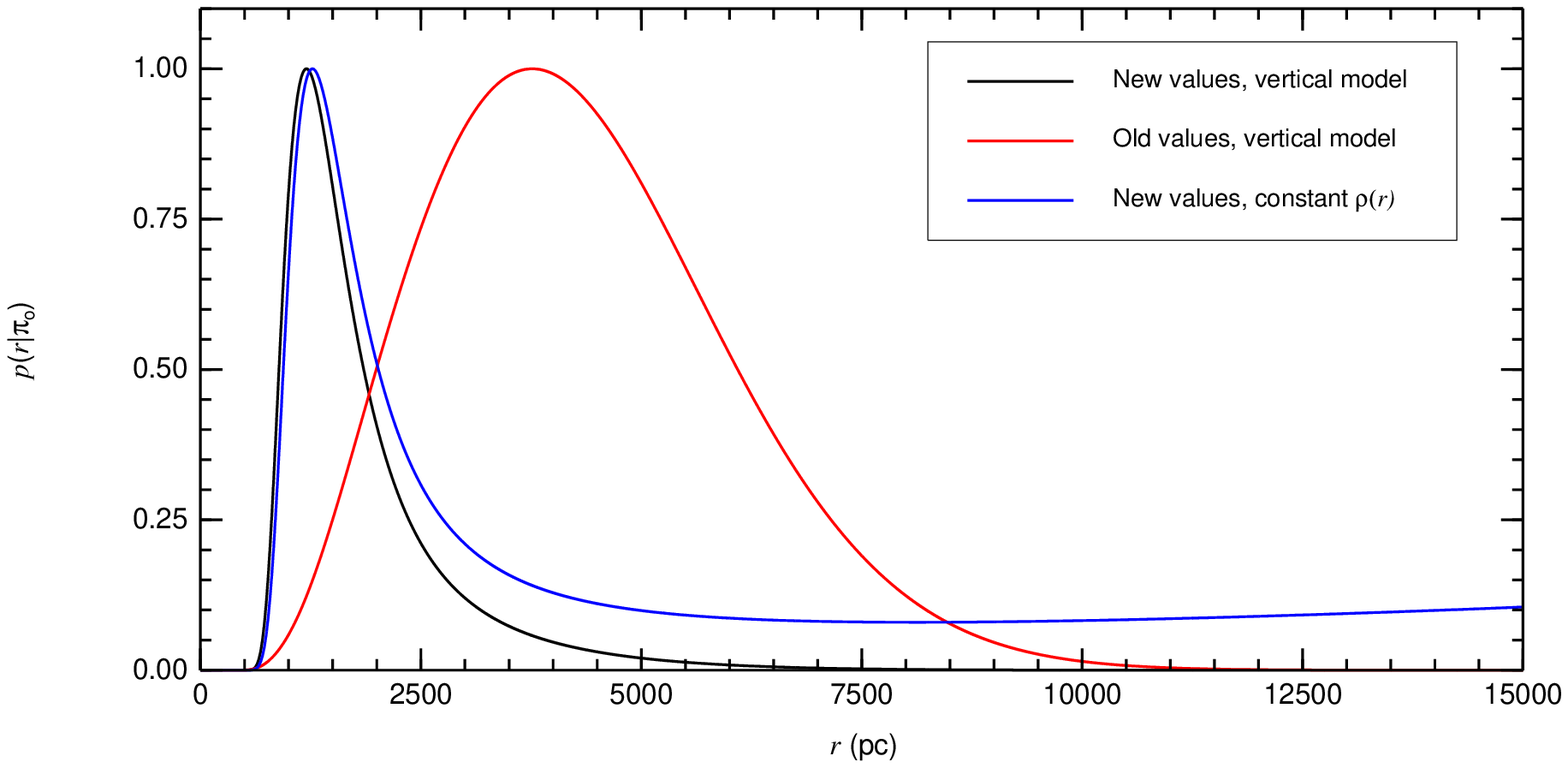}}

\noindent {\bf Figure 5.} $p(r|\pio)$ for HD 188209 for three cases: [a] New values of \pio, \spi, and $\rho(r)$, [b] old
values of \pio, \spi, and $\rho(r)$, and [c] new values of \pio, and \spi\ with constant $\rho(r)$. In the three cases 
$p(r|\pio)$ has been normalized to its peak value in the plotted range for clarity.

\bigskip

\input{ostars}

\medskip

\noindent {\bf Table 1.} The twelve nearest O and WR stars with significant Hipparcos distances.

\bigskip

\input{brightstars}

\medskip

\noindent {\bf Table 2.} Results for massive stars with $V_{\rm T} \le 1.5$.

\bigskip

\section*{References}

\begin{itemize}
  \item Elías, F., Cabrera Caño, J., \& Alfaro, E. J. 2006, {\it AJ} {\bf 131}, 2700
  \item	Lutz, T. E. \& Kelker, D. H. 1973, {\it PASP} {\bf 85}, 573
  \item	Maíz Apellániz, J. 2001, {\it AJ} {\bf 121}, 2737
  \item	Maíz Apellániz, J. 2005, {\it The Three-Dimensional Universe with Gaia}, 179
  \item	Maíz Apellániz, J., Walborn, N. R. et al. 2004, {\it ApJS} {\bf 151}, 103
  \item	Maíz Apellániz, J. \& \'Ubeda, L. 2005, {\it ApJ} {\bf 629}, 873
  \item	Menten, K. M. et al. 2007, {\it A\&A} {\bf 474}, 515
  \item	Millour, F. et al. 2007, {\it A\&A} {\bf 464}, 107
  \item	North, J. R. et al. 2007, {\it MNRAS} {\bf 377}, 415
  \item	van der Hucht, K. A. 2001, {\it NewAstRev} {\bf 45}, 135
  \item	van Leeuwen F. 2007, {\it Hipparcos, the New Reduction of the Raw Data}, Springer
\end{itemize}

\end{document}

%% file: jesus.tex
\def\ind{$\,\!$\indent}
\def\p/{\mbox{$^1$}}
\def\pp/{\mbox{$^2$}}
\def\ppp/{\mbox{$^3$}}
\def\pppp/{\mbox{$^4$}}
\def\m/{\mbox{$^{-1}$}}
\def\mm/{\mbox{$^{-2}$}}
\def\mmm/{\mbox{$^{-3}$}}
\def\mmmm/{\mbox{$^{-4}$}}
\def\C/{\mbox{\degr C}}
\def\K/{\mbox{\degr K}}
\newcommand{\vsmsp}[3]{\mbox{$#1_{- #2}^{+ #3}$}}
\newcommand{\thead}[1]{\multicolumn{1}{c}{#1}}
\def\ooo/{\mbox{$^{\underline{o}}$}}
\def\aaa/{\mbox{$^{\underline{a}}$}}
\def\xm/{\mbox{$\overline{x}$}}
\def\ym/{\mbox{$\overline{y}$}}
\def\x2m/{\mbox{$\overline{x^2}$}}
\def\y2m/{\mbox{$\overline{y^2}$}}
\def\xym/{\mbox{$\overline{xy}$}}
\def\Ms/{\mbox{M$_{\odot}$}}
\def\Rs/{\mbox{R$_{\odot}$}}
\def\Ts/{\mbox{T$_{\odot}$}}
\def\Ls/{\mbox{L$_{\odot}$}}
\def\Mt/{\mbox{M$_{\oplus}$}}
\def\Rt/{\mbox{R$_{\oplus}$}}
\def\Tt/{\mbox{T$_{\oplus}$}}
\def\Lt/{\mbox{L$_{\oplus}$}}
\def\mv/{\mbox{m$_{\rm v}$}}
\def\Mv/{\mbox{M$_{\rm v}$}}
\def\dias/{\mbox{$^{\rm d}$}}
\def\horas/{\mbox{$^{\rm h}$}}
\def\minutos/{\mbox{$^{\rm m}$}}
\def\segundos/{\mbox{$^{\rm s}$}}
\def\oiii/{\mbox{O\,{\sc iii}}}
\def\hi/{\mbox{H\,{\sc i}}}
\def\hii/{\mbox{H\,{\sc ii}}}

%% file: ostars.tex
\centerline{
\setlength{\tabcolsep}{6pt}
\begin{tabular}{lrlrrlr}
\hline
Name             & \thead{Old \pio} & \thead{Old $d$}           & \thead{Old $c$} & \thead{New \pio} & \thead{New $d$}       & \thead{New $c$} \\
                 & \thead{(mas)}    & \thead{(pc)}              & \thead{(pc)}    & \thead{(mas)}    & \thead{(pc)}          & \thead{(pc)}    \\
\hline
$\zeta$ Oph      & $7.12\pm 0.71$   & \vsmsp{144}{14}{17}       &           $4.0$ & $8.92\pm 0.21$   & \vsmsp{112}{3}{3}     &           $0.2$ \\
$\delta$ Ori A   & $3.56\pm 0.83$   & \vsmsp{323}{70}{152}      &          $42.3$ & $4.72\pm 0.58$   & \vsmsp{221}{25}{33}   &           $9.5$ \\
$\zeta$ Ori A    & $3.99\pm 0.79$   & \vsmsp{278}{50}{79}       &          $27.5$ & $4.44\pm 0.64$   & \vsmsp{239}{32}{43}   &          $14.1$ \\
15 Mon           & $3.19\pm 0.73$   & \vsmsp{514}{180}{6813}    &         $200.3$ & $3.54\pm 0.50$   & \vsmsp{309}{43}{60}   &          $26.2$ \\
$\zeta$ Pup      & $2.33\pm 0.51$   & \vsmsp{545}{126}{243}     &         $115.3$ & $3.00\pm 0.10$   & \vsmsp{335}{11}{12}   &           $1.5$ \\
$\gamma^2$ Vel   & $3.88\pm 0.53$   & \vsmsp{278}{37}{50}       &          $20.5$ & $2.99\pm 0.32$   & \vsmsp{349}{35}{44}   &          $14.4$ \\
$\lambda$ Ori A  & $3.09\pm 0.78$   & \vsmsp{405}{102}{293}     &          $80.9$ & $3.03\pm 0.55$   & \vsmsp{361}{60}{89}   &          $31.5$ \\
$\mu$ Col        & $2.52\pm 0.55$   & \vsmsp{486}{121}{253}     &          $89.5$ & $2.46\pm 0.20$   & \vsmsp{412}{32}{38}   &           $5.4$ \\
$\xi$ Per        & $1.84\pm 0.70$   & \vsmsp{1913}{1255}{1801}  &        $1369.8$ & $2.61\pm 0.52$   & \vsmsp{416}{74}{116}  &          $33.3$ \\
HD 149404        & $1.07\pm 0.89$   & \vsmsp{11801}{5301}{4934} &       $10866.3$ & $2.40\pm 0.36$   & \vsmsp{458}{67}{96}   &          $40.9$ \\
$\theta^2$ Ori A & $1.72\pm 1.00$   & \vsmsp{1867}{887}{1082}   &        $1285.3$ & $2.11\pm 0.42$   & \vsmsp{520}{103}{201} &          $46.2$ \\
10 Lac           & $3.08\pm 0.62$   & \vsmsp{353}{64}{107}      &          $28.1$ & $1.88\pm 0.22$   & \vsmsp{542}{59}{77}   &          $10.5$ \\
\hline
\end{tabular}
}

%% file: brightstars.tex
\centerline{
\setlength{\tabcolsep}{6pt}
\begin{tabular}{lrlrrlr}
\hline
Name                & \thead{Old \pio} & \thead{Old $d$}           & \thead{Old $c$} & \thead{New \pio} & \thead{New $d$}       & \thead{New $c$} \\
                    & \thead{(mas)}    & \thead{(pc)}              & \thead{(pc)}    & \thead{(mas)}    & \thead{(pc)}          & \thead{(pc)}    \\
\hline
Canopus             & $10.43\pm 0.53$  & \vsmsp{97}{5}{5}          &           $0.9$ & $10.56\pm 0.56$  & \vsmsp{96}{5}{5}      &           $1.0$ \\
Rigel               &  $4.22\pm 0.81$  & \vsmsp{254}{44}{71}       &          $17.2$ &  $3.78\pm 0.34$  & \vsmsp{267}{22}{27}   &           $2.9$ \\
Achernar            & $22.68\pm 0.57$  & \vsmsp{44}{1}{1}          &           $0.1$ & $23.38\pm 0.57$  & \vsmsp{43}{1}{1}      &           $0.1$ \\
$\beta$ Cen         &  $6.21\pm 0.56$  & \vsmsp{167}{14}{18}       &           $5.6$ &  $8.32\pm 0.50$  & \vsmsp{122}{7}{8}     &           $1.8$ \\
$\alpha$ Cru        & $10.17\pm 0.67$  & \vsmsp{100}{6}{7}         &           $1.8$ & $10.07\pm 0.48$  & \vsmsp{100}{5}{5}     &           $0.9$ \\
Betelgeuse          &  $7.63\pm 1.64$  & \vsmsp{176}{44}{105}      &          $44.9$ &  $6.56\pm 0.83$  & \vsmsp{164}{20}{27}   &          $11.2$ \\
Spica               & $12.44\pm 0.86$  & \vsmsp{81}{5}{6}          &           $1.0$ & $13.07\pm 0.69$  & \vsmsp{77}{4}{4}      &           $0.6$ \\
$\beta$ Cru         &  $9.25\pm 0.61$  & \vsmsp{110}{7}{8}         &           $1.9$ & $11.70\pm 0.98$  & \vsmsp{88}{7}{8}      &           $2.5$ \\
Antares             &  $5.40\pm 1.68$  & \vsmsp{1027}{799}{2002}   &         $841.6$ &  $5.90\pm 1.00$  & \vsmsp{187}{30}{44}   &          $17.2$ \\
Deneb               &  $1.01\pm 0.57$  & \vsmsp{13225}{6975}{4681} &       $12234.8$ &  $2.29\pm 0.32$  & \vsmsp{475}{65}{90}   &          $38.3$ \\
$\varepsilon$ CMa A &  $7.57\pm 0.57$  & \vsmsp{135}{10}{11}       &           $3.0$ &  $8.06\pm 0.14$  & \vsmsp{124}{2}{2}     &           $0.1$ \\
\hline
\end{tabular}
}